\documentclass[sigconf,preprint]{acmart}
\usepackage{enumitem}
\usepackage{graphicx}
\usepackage{colortbl}
\usepackage{multirow}
\usepackage{xcolor}
\usepackage{tcolorbox} 
\usepackage{booktabs}
\usepackage{subfig}
\usepackage{lipsum}
\usepackage{algorithm}
\usepackage{algorithmic}
\AtBeginDocument{%
  }

\definecolor{blue-violet}{rgb}{0.54, 0.17, 0.89}
\newcommand{\zjz}[1]{\textcolor{black}{#1}}
\newcommand{\bkq}[1]{{\color{black}{#1}}}

\acmConference[Conference acronym 'XX]{Make sure to enter the correct
  conference title from your rights confirmation emai}{June 03--05,
  2018}{Woodstock, NY}

\setcopyright{acmlicensed}
\copyrightyear{2018}
\acmYear{2018}
\acmDOI{XXXXXXX.XXXXXXX}
\acmISBN{978-1-4503-XXXX-X/18/06}

\begin{document}

\title{Real-Time Personalization for LLM-based Recommendation with Customized In-Context Learning }

\author{Keqin Bao*}
\email{baokq@mail.ustc.edu.cn}
\affiliation{%
  \institution{University of Science and Technology of China}
  \city{Hefei}
  \country{China}
}
\author{Ming Yan*}
\email{ym689@mail.ustc.edu.cn}
\affiliation{%
  \institution{University of Science and Technology of China}
  \city{Hefei}
  \country{China}
}

\author{Yang Zhang}
\email{zy1580@gmail.com}
\affiliation{%
  \institution{National University of Singapore}
  \city{Kent Ridge}
  \country{Singapore}
}
\author{Jizhi Zhang}
\email{cdzhangjizhi@mail.ustc.edu.cn}
\affiliation{%
  \institution{University of Science and Technology of China}
  \city{Hefei}
  \country{China}
}
\author{Wenjie Wang}
\email{wenjiewang96@gmail.com}
\affiliation{%
  \institution{National University of Singapore}
  \city{Kent Ridge}
  \country{Singapore}
}

\author{Fuli Feng}
\email{fulifeng93@gmail.com}
\affiliation{%
  \institution{University of Science and Technology of China}
  \city{Hefei}
  \country{China}
}
\author{Xiangnan He}
\email{xiangnanhe@gmail.com}
\affiliation{%
  \institution{University of Science and Technology of China}
  \city{Hefei}
  \country{China}
}

\thanks{* denotes equal contribution}

%



\begin{abstract}

 Frequently updating Large Language Model (LLM)--based recommender systems to adapt to new user interests --- as done for traditional ones --- is impractical due to high training costs, 
 even with acceleration methods. This work explores adapting to dynamic user interests without any model updates by leveraging \textit{In-Context Learning} (ICL), which allows LLMs to learn new tasks from few-shot examples provided in the input. Using new-interest examples as the ICL few-shot examples, LLMs may learn real-time interest directly, avoiding the need for model updates. However, existing LLM-based recommenders often lose the in-context learning ability during recommendation tuning, while the original LLM's in-context learning lacks recommendation-specific focus. 
 To address this, we propose RecICL, which customizes recommendation-specific in-context learning for real-time recommendations. RecICL organizes training examples in an in-context learning format, ensuring that in-context learning ability is preserved and aligned with the recommendation task during tuning. 
 Extensive experiments demonstrate RecICL's effectiveness in delivering real-time recommendations without requiring model updates. Our code is available at \url{https://github.com/ym689/rec_icl}.
\end{abstract}
\maketitle
\section{Introduction}

In recent years, Large Language Model (LLM)-based recommendation has emerged as a rapidly evolving field, demonstrating substantial potential to transform recommendations across diverse scenarios~\cite{llmrecsurvey1,llmrecsurvey2,harte2023leveraging}. 
Substantial efforts have been dedicated to this area, creating various mechanisms to align LLM capabilities with recommendation tasks.
Among these mechanisms, fine-tuning LLMs on recommendation data via instruction tuning has emerged as a popular approach, as it can \bkq{bridge the inherent absence of recommendation tasks in LLM pretraining}~\cite{recommendation_instruction, TALLRec, bao2023bi,lcrec}.
Numerous studies have investigated this approach, showcasing highly promising progress.

While significant progress has been made, exciting developments have overlooked addressing dynamic interests, a crucial aspect for real-world applications. In the real world, user interest can shift rapidly due to the varying instant interest in dynamic environments~\cite{sRec,ICF,streamrec,streambg3}. Traditionally, updating the model periodically with incoming data is commonly used to capture timely user interests~\cite{streamrec,streambg1,streambg2}. 
However, in the case of LLMs, updates at the model level incur substantial computational and time costs due to their massive number of parameters.
This limits the application of the periodical update approach. 
Although various acceleration methods can expedite the update process, the remaining costs are still excessively high compared to the timeliness requirements of recommender systems.
Recognizing the persistent issue of high updating costs for LLMs, this study explores the possibility of adapting the model to dynamic user interests without any model-level updates.

Among existing techniques, \textit{In-Context Learning} (ICL)  seems to be a promising choice for timely learning user drift interests without model updates. ICL is a well-known mechanism for LLMs,  by which LLMs can learn tasks by simply being given few-shot task examples in their input context, without any updates to their internal parameters~\cite{brown2020language, zhao2023survey}. It has demonstrated great success for cross-task generalization in the Natural Language Processing (NLP) domain. By incorporating recent user interaction data as few-shot task examples in the context, we can expect that ICL could effectively capture the interest drift from these examples. This process is carried out entirely at inference time, allowing for adapting to user real-time interests without necessitating model updates.

However, we argue that directly applying ICL to LLM-based recommendations is not feasible. Intuitively, there seem to be two directed approaches: 1) employing ICL with general LLMs and 2) employing ICL with LLMs specifically tuned for recommendations. Both approaches encounter challenges --- 
the first lacks alignment with recommendation tasks, while the second often suffers from diminished ICL capabilities for the tuned LLMs.
These challenges would undoubtedly hinder the practical application of these methods.
To effectively leverage ICL for real-time recommendations, we need a method that possesses recommendation-specific ICL capabilities --- aligning the model with recommendation tasks while preserving and even enhancing its ICL abilities.

To achieve this, we propose \textbf{\textit{RecICL}}, a method designed to customize recommendation-specific ICL in LLMs for real-time recommendations. RecICL optimizes the recommendation instruction tuning phase to preserve ICL capabilities. Specifically, when training the model, instead of only providing historical interaction sequences as input, we also include \bkq{recent recommendation examples (i.e., pairs of historical interactions and the next interaction)} as the context within the input. 
This allows the model to be fine-tuned in a way that naturally maintains its in-context learning ability, teaching it to leverage contextual information for recommendations. During inference, we replace these examples with real-time user data, allowing the model to adapt to evolving user interests in a few-shot manner. Extensive experiments on real-world datasets demonstrate the effectiveness of RecICL in making real-time recommendations for LLM-based recommendation.   

\begin{itemize}[leftmargin=*]
    \item To our knowledge, we are the first to explore adapting LLM-based recommenders to dynamic user interests without requiring any model-level updates.
    
    \item We propose RecICL, a novel approach for tailoring recommendation-specific in-context learning in LLMs to enable real-time recommendations.
    
    \item Experimental results demonstrate that our method significantly outperforms existing approaches and can also maintain robust performance over extended periods.
    
\end{itemize}

\section{Related Work}
In this section, we will introduce two topics related to our paper -- Streaming Recommender System and LLM-based Recommender System.

\subsection{Streaming Recommender System}
Streaming recommender systems have gained significant attention in the past years due to their ability to handle dynamic user interest and item catalogs in real-time scenarios~\cite{streambg1,streambg2,streambg3}. 
Unlike traditional batch-based recommender systems which only train and test on static fixed datasets, streaming approaches can efficiently process continuous streams of data and provide up-to-date recommendations~\cite{streamrec, ICF, IMF}.
Early work in this area focused on adapting traditional collaborative filtering techniques to streaming environments. For instance, researchers proposed an incremental collaborative filtering algorithm that updates user similarities on the fly as new ratings arrive~\cite{ICF}, which is more suitable for online applications than traditional collaborative filtering. Similarly, people introduced a streaming matrix factorization method that incrementally updates latent factors for users and items~\cite{IMF}. More recently, given to the success of Graph Neural Networks in recommender systems~\cite{lightGCN,graphrecsurvey}, researchers tend to apply continual graph learning techniques for streaming recommender systems~\cite{streamgraph1,streamgraph2, streamgraph3, streamgraph4}. In detail, recent work proposes a continual learning-based GNN model that can efficiently detect new patterns, update node representations, and consolidate existing knowledge while learning incrementally from streaming network data~\cite {streambg1}. To further address the issue of catastrophic forgetting in incremental graph learning, researchers have integrated two approaches: data replay and model regularization~\cite{streambg2}. 
Moreover, the DECG~\cite{streamgraph4} framework has been proposed to distinguish between outdated short-term preferences from useful long-term preferences, retaining only the beneficial long-term preference parameters and extracting new short-term preferences. All of those methodologies have achieved significant success in the streaming recommendation field. However, they primarily rely on timely updates and iterations of the model parameters, which can be challenging for LLMs due to their high cost.
In this paper, we draw inspiration from these previous studies and propose to tackle the challenge of streaming recommendations in the domain of LLMs for recommendation. 

%

\subsection{LLM-based Recommender System} 
%
Recently, inspired by the powerful and comprehensive capabilities of LLMs, an increasing number of researchers have been exploring various ways to leverage LLMs for recommendation~\cite{llmrecsurvey1, llmrecsurvey2, tutorial1, llmrec_survey_3}. Some researchers have attempted to effectively transfer the knowledge and capabilities of LLMs to traditional recommendation models~\cite{liu2024large, cui2024distillation}. They have experimented with extracting embedding information for text description, such as item titles and recommendation rationales~\cite{encoder2,encoder1,wei2024llmrec}. Then feeding this embedding information into conventional recommendation models to make predictions. While these methods can address the issue of user interest drift through the iterative process of traditional models, they face two significant challenges. First, existing decoder models are not well-suited for extracting embeddings, which greatly limits the potential of LLMs. Second, such methods often require fine-tuning LLMs on static data to optimize embeddings, which can significantly impact the model's representational capabilities when data is updated or user interests change.

Another group of people focuses on harnessing the generative power of LLMs to produce recommendations directly~\cite{lin2023multi, liao2023llara, recmind, zhang2023collm, IDGenRec, yang2023palr, he2023large}. 
In detail, those researchers have noted that LLMs are exposed to limited recommendation data during their training phase, necessitating an alignment approach to learning recommendation tasks. Initially, researchers directly fine-tuned these models by organizing recommendation data into instruction-input-output formats. More recently, they have been continuously exploring ways to inject collaborative information learned from traditional models into LLMs during the training process to enhance their recommendation capabilities.
Although achieving great success, these methods focus on static and fixed datasets where user interest is stable. However, in real-world scenarios, data is constantly updated, user interests are changing, and new user feedback is continually generated~\cite{sRec,ICF,streambg3}. Therefore, in this paper, we will delve into discussing how LLM recommendations perform in scenarios with user interest drift and explore methods to mitigate this issue by utilizing newly generated user feedback.


\begin{figure}[t]
    \centering
    \begin{minipage}[t]{0.5\linewidth}
        \centering
        \includegraphics[width=\textwidth]{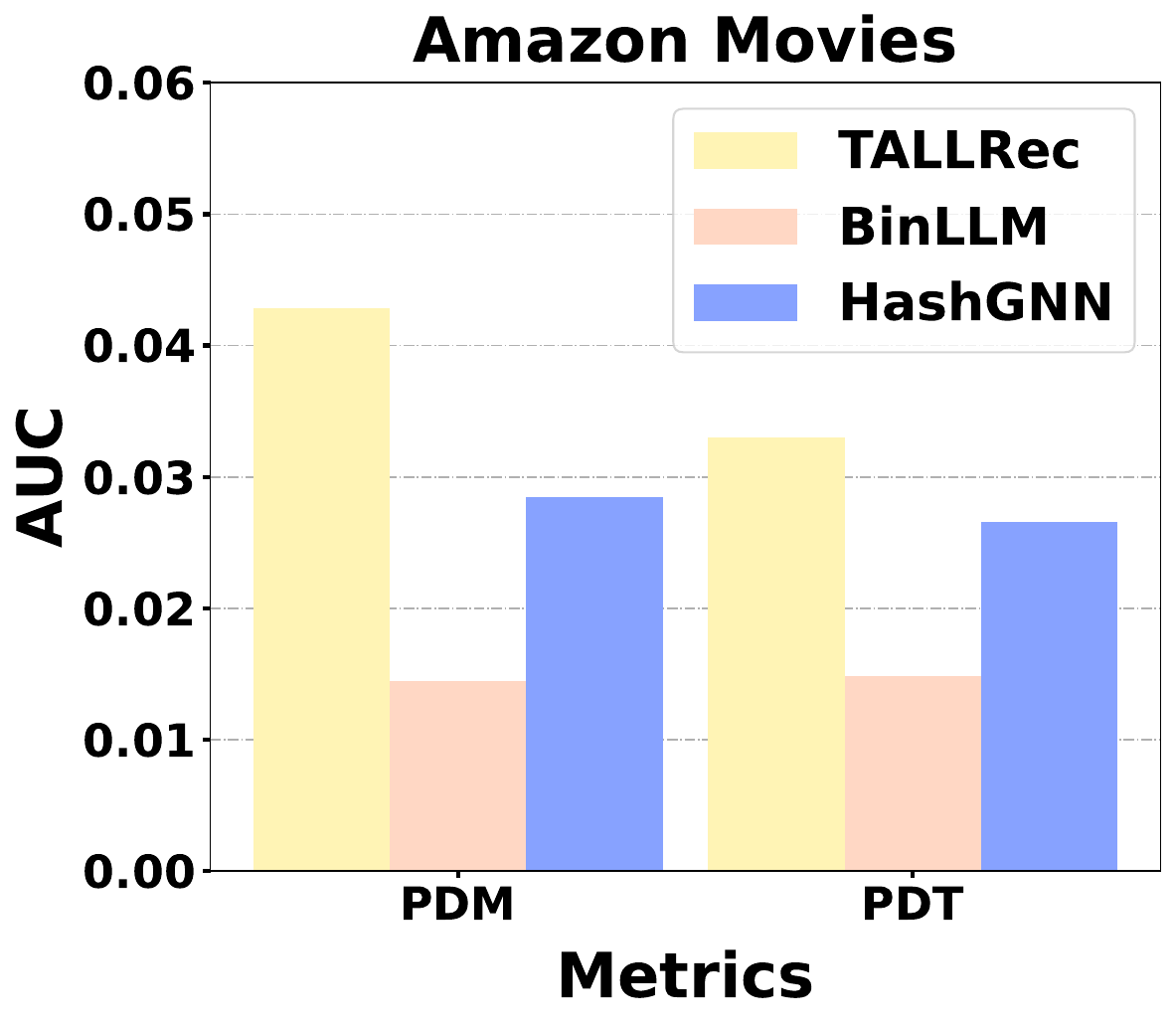}
        
    \end{minipage}%
    \begin{minipage}[t]{0.5\linewidth}
        \centering
        \includegraphics[width=\textwidth]{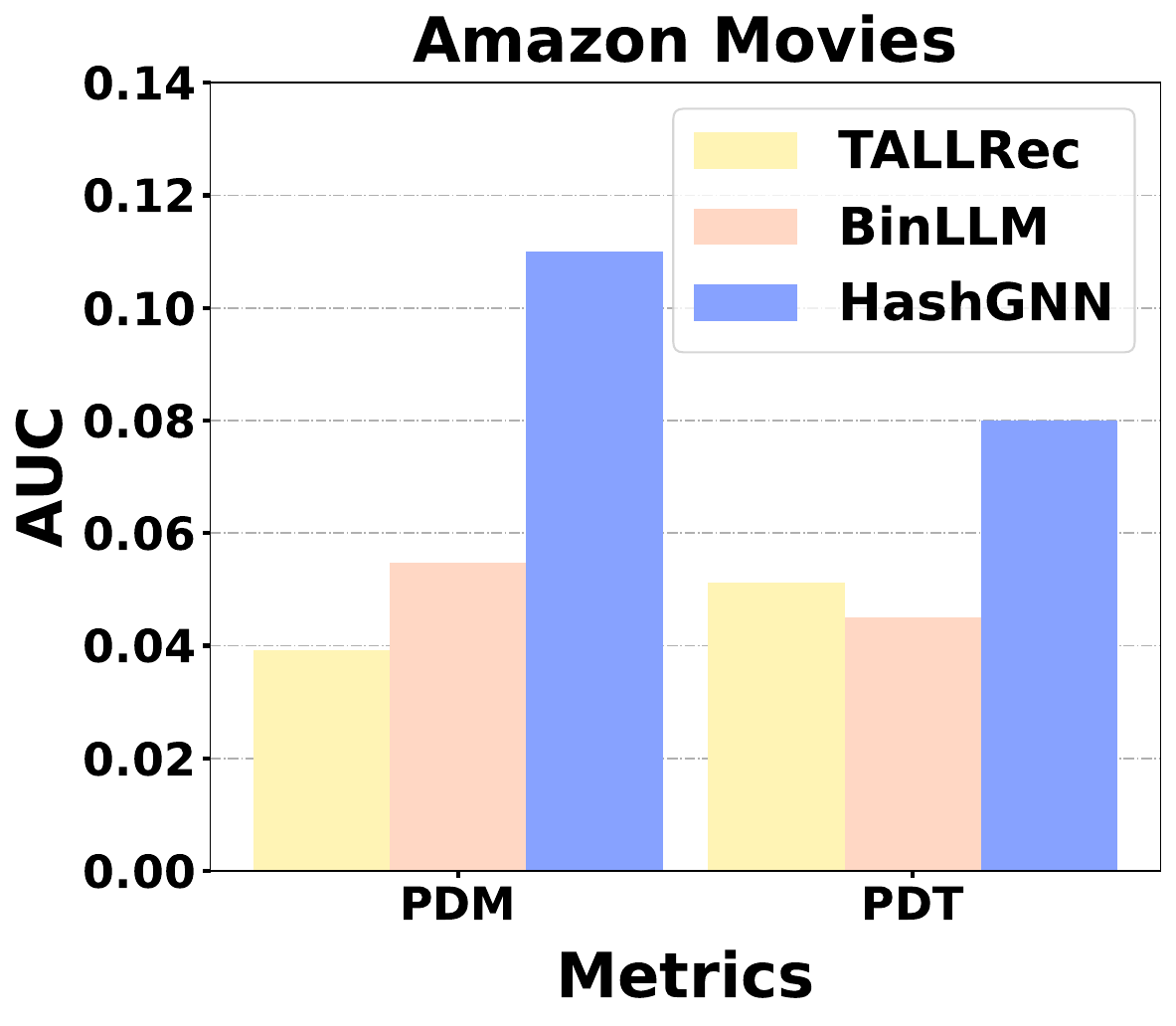}
   
    \end{minipage}
    \caption{A comparative analysis of TALLRec, BinLLM, and HashGNN models using both Amazon-Book and Amazon-Movie datasets for performance assessment.  A higher ``\zjz{PDM}'' indicates a greater benefit of updating the mode. A higher ``\zjz{PDT}'' signifies a more significant impact of shifts in user interests on the model's performance.}
    \label{fig:decline}  
\end{figure}

\section{Preliminary}

In this section, we first present the problem formulation (\S~\ref{sec:3.1}) for the studied user interest drift problem. Next, we present preliminary studies (\S~\ref{sec:3.2}) to illustrate the importance of timely adapting LLM-based recommenders to new user interests.


\subsection{Problem Defination}
\label{sec:3.1}
In recommendation, users are expected to continuously interact with the system, making the interaction data steamingly coming. We represent the streaming data $T$ as $\{D_0, \dots, D_t, \dots, \}$, where $D_t$ denotes the data collected at $t$-time period.   
During the process, their interest could evolve. To ensure the recommendation performance, we usually update the recommenders using the newly collected data to capture the new interest. However, for LLM-based recommender models, it is costly to update them. Therefore, we usually need to train a model with $\{D_0, \dots, D_T\}$  (denoting the trained model as $f_T$),
but need it to serve for many periods, e.g., $K$ periods, form $D_{T+1}$ to  $D_{T+K}$.
In this work, we consider developing a method that could capture the user's new interest (real-time interest) from the new data \textbf{without requiring model updates}.



\subsection{Importance of Adapting to New Interests}
\label{sec:3.2}
We conduct preliminary experiments using two representative LLM-based models, TALLRec and BinLLM, to verify the importance of adapting LLM-based recommenders to users' evolving interests. Our analysis is based on Amazon-Books and Amazon-Movies datasets. Specifically, we split the data into 10 periods and we use $\{D_0, \dots, D_4\}$ to train TALLRec and BinLLM, obtaining the trained model $f_4$. 
Additionally, we train the model on $\{D_0, \dots, D_8\}$ to obtain more updated model, $f_8$. We then compare the performance of the models on $D_5$ and $D_9$.
We define two metrics according to the performance difference between models to demonstrate the importance of capturing users' new interests:

\begin{itemize}[leftmargin=*]

\item \textit{$PDT$}: This metric measures the performance difference for the same model across two different testing period datasets. We compute PDT as:
  $$PDT = AUC(f_4; D_5) - AUC(f_4; D_9).$$ 
    \item \textit{$PDM$}: This metric measures the performance difference between a less updated model and a more updated model on the same testing data. We compute PDM as:
\begin{equation}
    \label{eq:pdm}
    PDM = AUC(f_8; D_9) - AUC(f_4; D_9),
\end{equation}
  where $AUC(f; D)$ represents the AUC evaluated for model $f$ on dataset $D$.

\end{itemize}


\textit{Results}. Figure~\ref{fig:decline} summarizes the results, with a traditional recommender model (HashGNN) included as a reference. From the figure, we can make the following observations: 1) Both TALLRec and BinLLM exhibit positive values on metric $PDT$, indicating that as the testing data shifts from $D_5$ to $D_9$, the model $f_4$ experiences a noticeable performance decline, similar to traditional models. 2) TALLRec and BinLLM also show positive values on metric $PDM$, meaning that the more updated model ($f_8$) outperforms the less updated model ($f_4$) when tested on $D_9$. This demonstrates that updating the model leads to improved results. The results of LLM-based recommenders are generally consistent with those of traditional models. All the results confirm that LLM-based recommenders also need to adapt to users' evolving interests; otherwise, they risk sub-optimal performance.

\begin{figure}[t]
    \begin{minipage}[t]{0.5\linewidth}
        \centering
        \includegraphics[width=\textwidth]{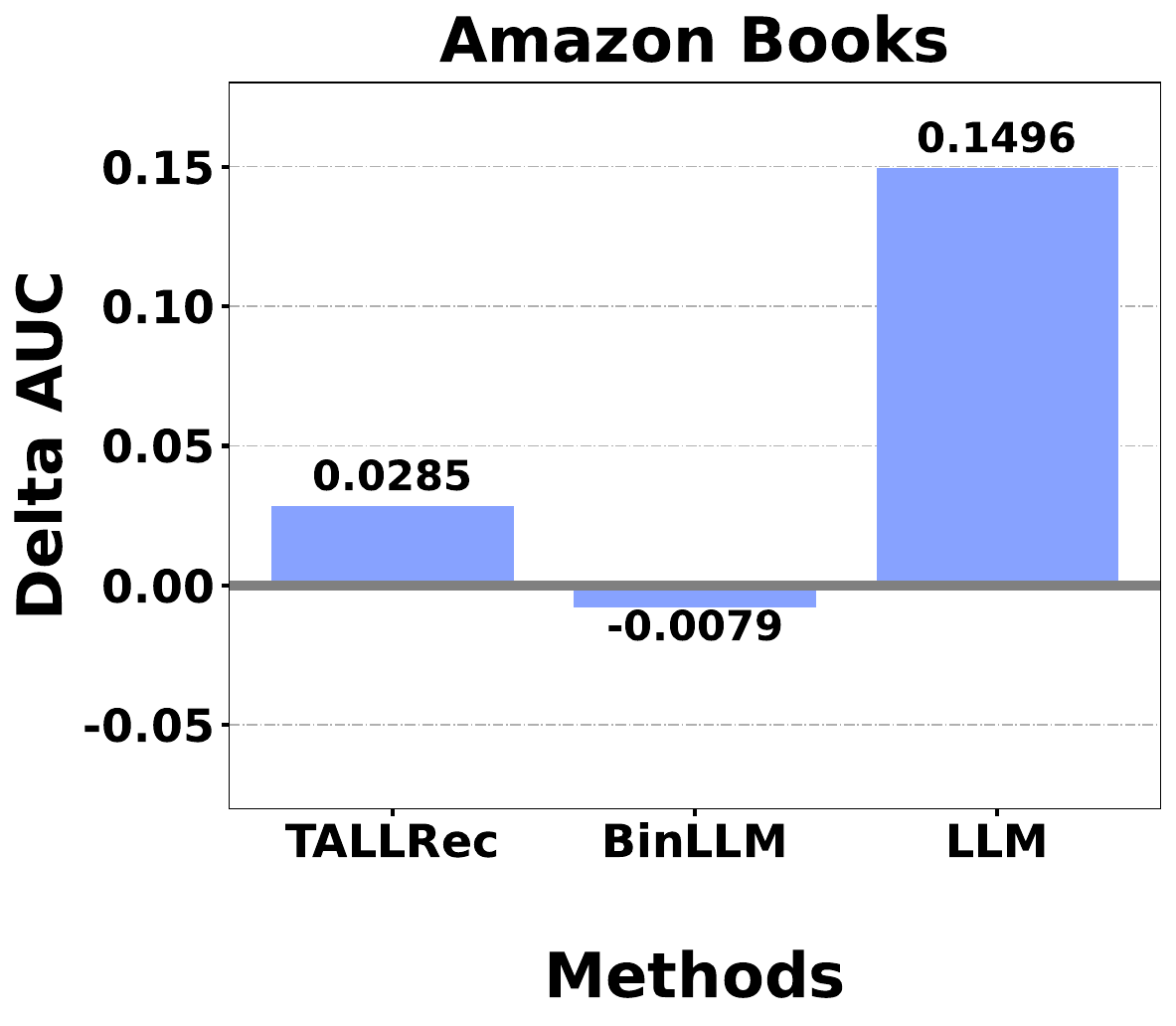}
        
    \end{minipage}%
    \begin{minipage}[t]{0.5\linewidth}
        \centering
        \includegraphics[width=\textwidth]{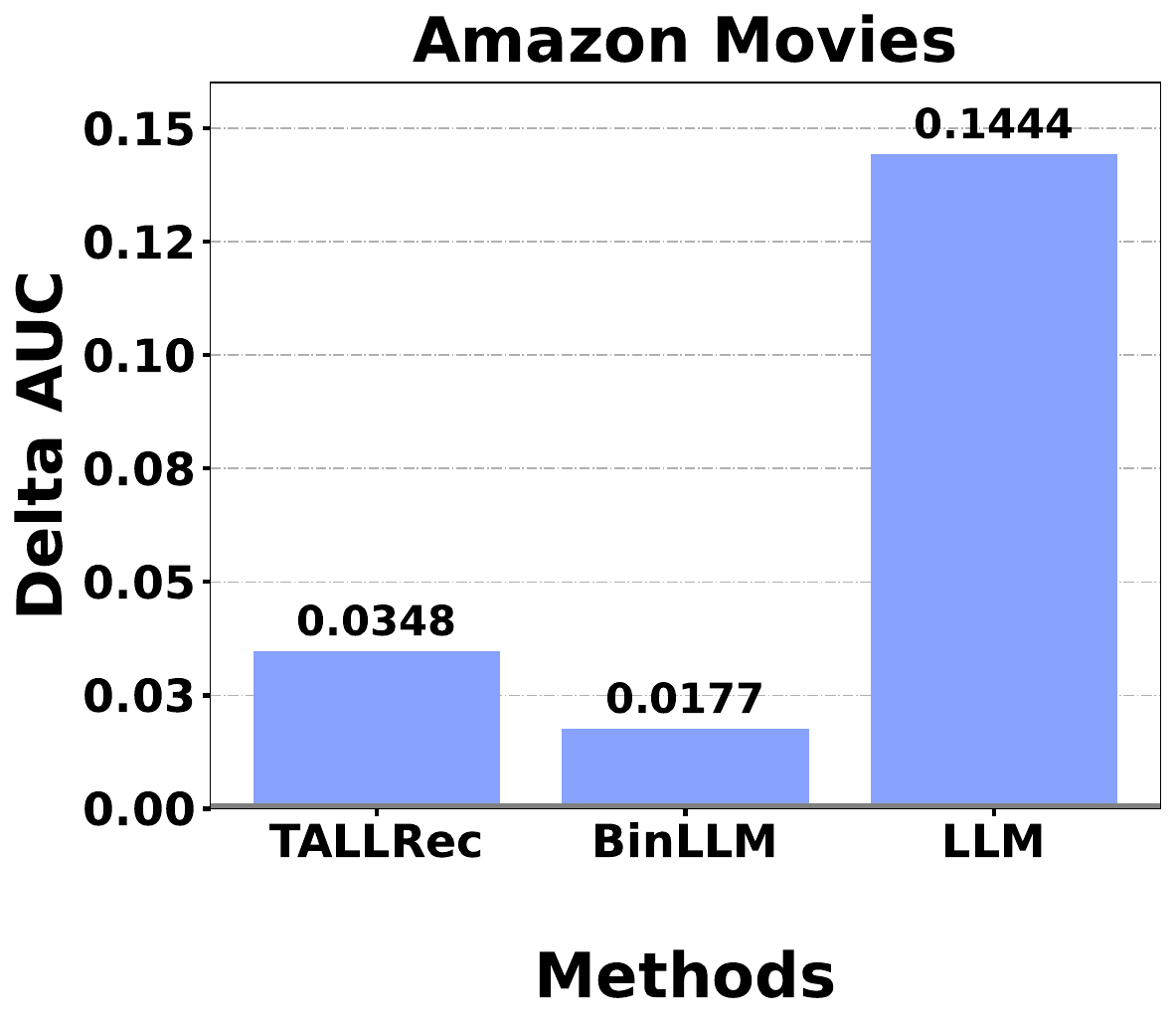}
   
    \end{minipage}
    \caption{Performance of TALLRec, BinLLM, and General LLM in scenarios with and without in-context learning on each dataset. The x-axis shows the name of the method, while the y-axis shows the performance improvement brought by using in-context learning, where in-context learning refers to selecting the four most recent interactions for each user to formulate the few-shot examples.}
    \label{fig:icl_delta}
\end{figure}
\begin{figure*}[htbp]
    \centering
    \includegraphics[width=0.85\textwidth]{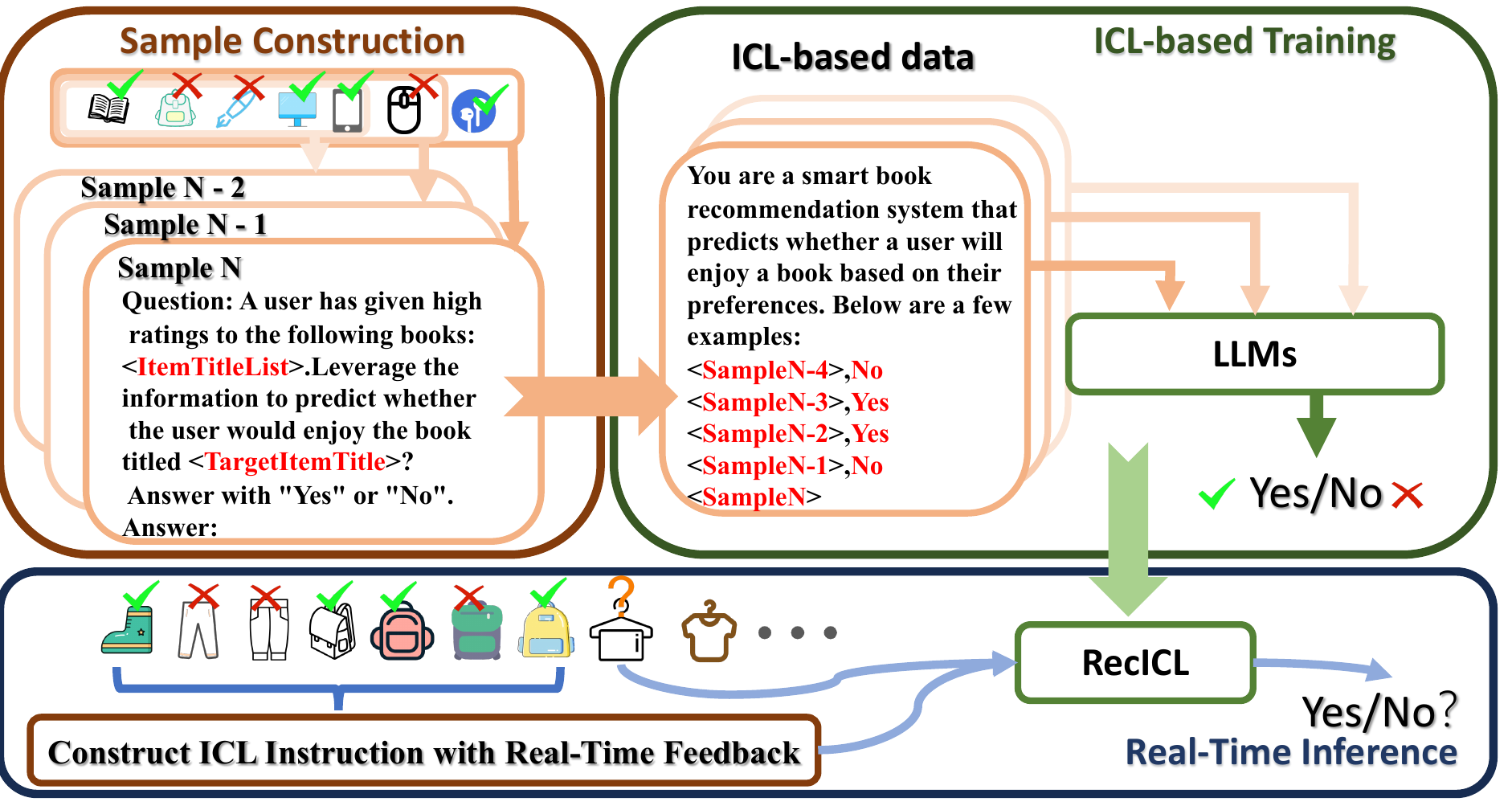}  
    \caption{Overview of our RecICL pipeline,  primarily consists of three stages: Sample Construction, Model Training, and Real-time Inference. Here we define the few-shot number as 4. }
    \label{fig:pipeline}  
\end{figure*}
\section{Method}




In this section, we first explore the potential and challenges of using \textit{In-Context Learning} (ICL) to adapt LLMs to users' evolving interests without requiring model updates. We then introduce the proposed \textbf{\textit{RecICL}} method, which is designed to customize recommendation-specific in-context learning for capturing real-time user interests.


\subsection{The Potential of ICL}
\label{sec:3.3}

In-context learning holds considerable promise for addressing user interest drift. As a recognized technology for LLMs, ICL allows the LLMs to rapidly adapt to new tasks using few-shot examples included in the input, without requiring any updates to the model parameters. 
In our scenarios, we can expect that integrating user feedback on new interests as few-shot examples will improve the model's capacity to capture such new interests for recommendation.

However, directly applying (ICL) to existing LLM-based recommender models may not be feasible, as these models can lose ICL capabilities during the tuning process. We compare the differences in ICL capabilities between these models and general LLMs by examining the performance improvements that ICL can provide. Specifically, following the setting in \S\ref{sec:3.2}, we divide the dataset ${D}$ into 10 periods and use the first five periods, $\{D_0, \dots, D_4\}$, to train BinLLM and TALLRec. For each model, we compare the performance (AUC) with and without ICL to determine the performance improvements brought by applying ICL, denoted the result as Delta AUC. Figure~\ref{fig:icl_delta} summarizes the results on  $D_9$. The findings indicate that TALLRec and BinLLM show minimal performance enhancement when utilizing ICL and may even experience a decline in performance compared to general LLMs. This suggests that LLM-based recommenders have lost their ICL capabilities. Furthermore, we cannot directly solve our problem using the ICL capabilities of general LLMs, as these models lack sufficient recommendation abilities. This motivates us to preserve the ICL abilities when equipping LLMs with recommendation capabilities.

\subsection{RecICL}
In this section, we present the proposed RecICL. We begin by providing an overview and then delve into each component in detail.

\subsubsection{Overview}
With the insight of keeping the ICL abilities when tuning LLMs to recommendation tasks,
we proposed RecICL for achieving real-time recommendation. The core lies in directly organizing the training example in an in-context learning format, making the ICL usable during tuning. 
As Figure~\ref{fig:pipeline} shows, our total RecICL framework for achieving real-time recommendations includes three main components:
\begin{itemize}[leftmargin=*]
    \item \textbf{Sample Construction}: In this component, we structure each interaction into an instruction format for next-step use, creating samples of prompts.
    
    \item \textbf{ICL-based Tuning}: 
    This component corresponds to the tuning process of RecICL. Rather than using traditional supervised fine-tuning (SFT), we design a new ICL-based tuning method to preserve the model’s ICL abilities while tuning the model on recommendations data. We organize the training samples in an ICL format by incorporating the most recent samples as few-shot examples, which are then used to fine-tune LLMs. As explicitly reinforced by the structure of the ICL-based training data, the ICL capabilities are expected to be retained during tuning.


    \item \textbf{Real-time Inference}:
    At inference, we also construct each testing sample in ICL format by feeding the real-time interactions as the few-shot examples. This ensures that the model can leverage ICL to learn the latest user interests to provide real-time recommendations.
\end{itemize}
We next elaborate on each component.



\subsection{Sample Construction}

In our framework, each sample would be formatted as an instruction, as shown on the left side of Figure~\ref{fig:pipeline}, where the task is described using language text. Let $(h_u, i, y)$ denote an interaction sample between a user $u$ and item $i$ in the dataset, where $y$ is the interaction label, and $h_u$ denotes the user's interaction history (including both interacted items and their labels). 
To convert each sample, we use the prompt template outlined in the “Sample Construction” part of Figure 3 --- feeding $h_u$ and $i$ into the `<ItemTitleList>' field and `<TargetItemTitle>' field in the prompt template, respectively. Let $x$ represent the prompt generated for the sample $(h_u,i,y)$; the new sample can be expressed as $(x, y)$.

For each user, we convert her/his all samples as above, obtaining a series of transformed samples:  
\begin{equation}  
    [(x_0, y_0), \dots, (x_n, y_n), \dots, (x_N, y_N)],  
\end{equation}  
where \((x_n, y_n)\) represents the \(n\)-th sample and \(N\) is the total number of samples for the user. 
Here, the samples are organized in chronological order based on the interaction timestamps. It is important to note that these samples are not directly used for model training; rather, they are prepared for constructing the ICL-based data.


\subsection{ICL-based Tuning}
To maintain the model’s ICL abilities while tuning it on recommendation data, we propose a new ICL-based tuning method. Instead of tuning the model to predict \(y_{n}\) based solely on \(x_{n}\) for each sample \((x_{n}, y_{n})\), we incorporate several of the most recent samples to help the predictions. These samples are integrated in an ICL format. Specifically, for a sample \((x_{n}, y_{n})\), we concatenate \(x_{n}\) with the \(M\) most recent samples (denoted as \(\{(x_{n-1},y_{n-1}), \dots, (x_{n-M}, y_{n-M})\}\)) as few-shot examples to construct the ICL instruction data, following the prompt template shown in the "ICL-based Tuning" section of Figure~\ref{fig:pipeline}. Let \(x_n^\prime\) represent the resulting prompt. Formally,
$$ 
x_n^\prime = P_{ICL}(\{(x_{n-1},y_{n-1}), \dots, (x_{n-M}, y_{n-M})\}; x_n),
$$
where \(P_{ICL}(\cdot)\) denotes the process for constructing the ICL instruction. Then the sample can be represented by $(x_{n}^{'},y_{n})$.

After generating all the ICL instruction data, we use it to tune the model, ensuring that it retains its ICL capabilities while learning the recommendation task. Specifically, we fine-tune the LLM by minimizing the following optimization objective:
\begin{equation}\label{eq:icltuning}
    \underset{\theta}{minimize} \sum_{(x_k^{\prime}, y_k)} \ell(f(x_k^{\prime}; \theta), y_k),
\end{equation}
where \(\theta\) represents the LLM’s parameters, \(f(x_k^{\prime}; \theta)\) denotes the model’s prediction for \(x_k^{\prime}\), and \(\ell\) is the commonly used cross-entropy loss in the context of LLMs.



By adopting this approach, we can achieve two goals simultaneously. On one hand, we can align the LLM with recommendation scenarios by training it on user interaction data. On the other hand, we can prevent the LLM from experiencing catastrophic forgetting of in-context learning ability, which occurs when it is trained on only a single sample. More importantly, this method teaches the model how to leverage the users' most recent interests from the few-shot examples during the training process,  enabling it to capture users' real-time interests at the inference stage.

\subsection{Real-time Inference}

During inference, the tuned LLM, with its ICL abilities preserved, can capture a user's new interests without requiring model updates. For each test sample, we simply use the real-time user feedback data as the few-shot example in ICL, allowing the model to access the user's newest interests. The process of constructing ICL instruction data for inference is identical to that used in training. Let \(x^{\prime}\) represent a test sample represented in ICL instruction format; the final prediction is made as  
$$\hat{y} = f(x^{\prime}, \theta^*),$$  
where \(\theta^*\) are the LLM model parameters tuned according to Equation~\eqref{eq:icltuning}.


\section{Experiments}







In this section, we will introduce the experiment setting and answer the following research questions:
\begin{itemize}[leftmargin=*]
    \item \textbf{RQ1}: How does RecICL perform under the user interest drift setting, and does it have any advantages compared to models updated with the full datasets?
    \item \textbf{RQ2}:Whether RecICL performance more stable across all time periods when comparing with other baselines?
    \item \textbf{RQ3}: What's the effect of different parts of RecICL?
\end{itemize}
\subsection{Experimental Settings}
\subsubsection{Datasets}
\begin{table}[t]
\caption{Data statistics of the datasets.}
\label{tab:dataStatics}
\renewcommand\arraystretch{1.1}
\begin{tabular}{cccc}
\hline
Dataset&\#Interaction&\#User&\#Item
\\ \hline
Amazon-Books &775,635&22,127&34,076\\
Amazon-Movies &378,329&11,799&14,632\\\hline
\end{tabular}
\label{table:data_statis}
\end{table}
We conduct our experiments on the following two real-world datasets:
\begin{itemize}[leftmargin=*]
    \item \textbf{Amazon-Books} refers to the ``book'' subset of Amazon Review datasets \footnote{\url{https://cseweb.ucsd.edu/~jmcauley/datasets/amazon_v2/}}. This dataset consists of user reviews of book products from the Amazon platform between 1996 and 2018, with rating scores ranging from 1 to 5. We chose 4 as the threshold. Those with scores higher than 4 are labeled as ``Yes''; otherwise, they are labeled ``No''.
    \item \textbf{Amazon-Movies} refers to the ``movie'' subset of Amazon Review datasets. Similar to the Amazon-Book datasets, we choose the threshold as 4.
\end{itemize}

To better simulate real-world scenarios that prevent data leakage~\cite{data_Leakage_1} while modeling user interest drifts, we divided the dataset into 10 parts similar to \S\ref{sec:3.2}. Consistent with our setup in the preliminary experiments, by default, we use $D_{train}=[D_{0},D_{1},\ldots,D_{4}]$ as the training set. The last 5,000 samples from $D_{4}$ are separated to form the validation set and we randomly select 5,000 samples from \zjz{$D_9$} to serve as the test set for user interest drift. 
Specifically, for Amazon-Books, we preserved user interactions from the year 2017, and for Amazon-Movies, we preserved user interactions from the year 2014 to 2016.
Besides, following the setting in BinLLM~\cite{binllm}, we filtered out users and items with fewer than 20 interactions to ensure data quality.
The detailed data statistics are shown in Table~\ref{table:data_statis}
\subsubsection{Evaluation and Metrics} 

We use AUC, a common metric in recommender systems, and $PDM$, a metric we defined in Equation~\eqref{eq:pdm}, to evaluate our model's performance. AUC stands for the Area under the ROC Curve, which quantifies the overall prediction accuracy. As for $PDM$, similar to \S\ref{sec:3.2}, we define it as the model's performance $PDM$ between the fully updated model and the less updated model testing on $D_9$. While AUC shows the overall performance in the context of user interest drift, $PDM$ indicates how close we are to the upper bound of performance.
\begin{table*}[ht]
\centering
\caption{Overall performance comparison on the Amazon-Book and Amazon-Movie datasets on the $D_{9}$. $\uparrow$ indicates higher values are better, while $\downarrow$ indicates lower values are better. ``Rel Imp'' denotes the relative improvement of RecICL compared to baselines on the AUC metric.``RBR'' demonstrates the percentage of RecICL performance compared to the baseline performance on the $PDM$ metric. A lower RBR value indicates a greater improvement in our method. ``Collab.'' refers to the traditional collaborative methods. ``LLMRec (ICL)'' refers to the use of in-context learning with corresponding methods. Note that, ICL-LLM does not have PDM value since the general LLM is no need to update. The best performance for each metric are bolded.}
\begin{tabular}{cc|cccc|cccc}
\toprule
\multicolumn{2}{c|}{Dataset}                                                                    & \multicolumn{4}{c|}{\zjz{Amazon-Books}}                              & \multicolumn{4}{c}{\zjz{Amazon-Movies}} \\ \hline
\multicolumn{2}{c|}{Methods}                                                                    & AUC($\uparrow$)              &Rel Imp($\uparrow$)  & $PDM$($\downarrow$) &  RBR($\downarrow$)                          & AUC($\uparrow$)              &Rel Imp($\uparrow$)  & $PDM$($\downarrow$) &  RBR($\downarrow$)             \\ \hline
\multicolumn{1}{c|}{\multirow{3}{*}{Collab.}}                    & MF                           & 0.6193        &22.08\%     & 0.0882        & 0.0351     & 0.5901       &32.44\% & 0.1565            & 0.0364           \\
\multicolumn{1}{c|}{}                                            & SASRec                       & 0.5734        &26.67\%     & 0.1125        & 0.0276     & 0.6554       &25.91\% & 0.1029            & 0.0554                      \\
\multicolumn{1}{c|}{}                                            & HashGNN                      & 0.7396        &10.05\%     & 0.0285        & 0.1088     & 0.6628       &25.17\% & 0.1100            & 0.0518                   \\ 
\hline                                  
\multicolumn{1}{c|}{\multirow{3}{*}{LLMRec (ICL)}}               & ICL-LLM                      & 0.7133        &12.68\%     & -             & -          & 0.7434       &17.10\% & -                 & -                           \\
\multicolumn{1}{c|}{}                                            & ICL-TALLRec                  & 0.7290        &11.11\%     & 0.0214        & 0.1449     & 0.7763       &13.82\% & 0.0285            & 0.2000              \\
\multicolumn{1}{c|}{}                                            & ICL-BinLLM                   & 0.7708        &6.93\%      & 0.0053        & 0.5849     & 0.7835       &13.10\% & 0.0604           & 0.0944             
\\ \hline
\multicolumn{1}{c|}{\multirow{3}{*}{LLMRec}}                     & TALLRec                      & 0.7005        &13.96\%     & 0.0428        & 0.0724     & 0.7415       &17.30\% & 0.0392            & 0.1454                       \\
\multicolumn{1}{c|}{}                                            & BinLLM                       & 0.7787        &6.14\%     & 0.0145        & 0.2138     & 0.7658       &14.87\% & 0.0547            & 0.1042              \\
\multicolumn{1}{c|}{}                                            & BinLLM+                      & 0.6659        &17.42\%     & 0.1273        & 0.0244     & 0.7585       &15.60\% & 0.0620            & 0.0919              \\
\hline

\multicolumn{1}{c|}{\multirow{2}{*}{Ours}}                                        & \zjz{\textbf{RecICL-TALLRec}}     & \textbf{0.8401}   &-            & \textbf{0.0031}             & -                 & \textbf{0.9145}    &- &\textbf{0.0057}             & -                   \\
\multicolumn{1}{c|}{}                                            & RecICL-BinLLM                    & 0.8353        & -     & 0.0104           & -          & 0.9055         &-      & 0.0143         & -                   \\
\bottomrule
\end{tabular}
\label{table:main_result}
\end{table*}

\subsubsection{Baselines.} 
To fully investigate the performance of our RecICL, we mainly consider two types of baseline models, one is conventional recommender systems, and another is the recommender systems based on LLM. In detail, we select the following baselines:
\begin{itemize}[leftmargin=*]
    \item \textbf{MF}~\cite{mf} refers to  Matrix Factorization, which is a popular collaborative filtering technique, which works by decomposing the user-item interaction matrix and representing latent factors for users and items for predictions.
    \item \textbf{SASRec}~\cite{SASRec} refers to Self-Attentive Sequential Recommendation, which leverages the self-attention mechanism to capture long-term user preferences and item relationships, allowing it to model complex sequential patterns in user behavior.
    \item \textbf{HashGNN}~\cite{hashgnn} refers to Hashing with GNNs, which consists of a GNN encoder and a hash layer for encoding representations to hash codes. It can be viewed as a representation of the GNN-based method for collaborative filtering.
    \item \textbf{ICL} refers to how we directly apply the in-context learning techniques to prompt the LLM to determine whether the user will enjoy the item by giving the most recent interactions and the feedback of the user.
    \item \textbf{TALLRec}~\cite{TALLRec} is a representation of LLM-based recommender systems that directly use instruction-tuning to finetune the LLM on recommendation data and achieve moderate performance.
    \item \textbf{BinLLM}~\cite{binllm} is currently a state-of-the-art (SOTA) method for aligning LLMs with recommendation. It introduces collaborative information to LLMs by compressing the embedding from traditional recommender systems to 32-bit binary sequences and feeding it into the LLMs. 
    \item \textbf{BinLLM+} is a variant of the \zjz{SOTA method}
    BinLLM. During the inference time, we update the traditional recommendation model to check if updating the traditional models could help the BinLLM learn about new user interests. 
\end{itemize}

\subsubsection{Implementation Details}
Similar to BinLLM~\cite{binllm}, for traditional recommender systems, we employ Binary Cross-Entropy (BCE) as the optimization loss and use the Adam optimizer~\cite{adam}, unless otherwise specified by the original paper. For hyperparameter tuning, we explore the learning rate in $[1e{-2}, 1e{-3}, 1e{-4}]$ and tune the weight decay in the range of  $[1e{-2}, 1e{-3}, \ldots, 1e{-7}]$. For embedding size, we perform tuning within the range of $[16, 32, 64, 128, 256, \\ 512]$. For SASRec, we set the maximum length of historical interaction sequences according to the average user interaction count in the training data, as specified in the original paper.
For all LLM-based methods, we employ the AdamW optimizer and adjust the learning rate within the range of $[1e{-3}, 1e{-4}, 1e{-5}]$. We set up 200 to the warm-up steps in our training process. Regarding the input interaction sequence length, we follow the TALLRec~\cite {TALLRec} approach by setting the maximum sequence length to 10. For BinLLM, we utilize the optimal HashGNN and adapt it for binary sequence embedding. As for the backbone, we opt for Qwen1.5-0.5B~\cite{bai2023qwen}, considering its convenience and efficiency.

\subsection{Main Results (RQ1)}

Table~\ref{table:main_result} presents the overall performance of our method on two datasets following significant user interest shifts. From this table, we can draw the following conclusions:
\begin{itemize}[leftmargin=*]
    \item Compared to all other methods, RecICL demonstrates significant improvements in the AUC metric. This indicates by enhancing the model's recommendation-specific in-context learning capabilities, RecICL can substantially enhance model performance when user interest drifts. Moreover, when examining $PDM$ performance, we observe that the benefit of updating the RecICL model is minimal (0.0031 and 0.0057 for each dataset). This highlights the stability of the RecICL method, as it can maintain high performance over extended periods without model updates. 

    \item When comparing traditional recommender systems with LLM-based recommender systems, a notable performance gap in their ability to adapt to changing user interests. On average, traditional models demonstrate higher sensitivity to user interest drift, as evidenced by their $PDM$ metrics hovering around 0.1 (except the hashGNN model on the Amazon-book dataset). This phenomenon, on one hand, underscores the need for more frequent model updates in traditional systems. In contrast, it also highlights the robustness of LLM in resisting the effect of user interest, showing the potential necessity of applying LLMs to overcome this problem.
    \item The comparative analysis of traditional recommender systems and LLM-based recommender systems reveals a significant disparity in their adaptability to evolving user interests. Traditional models exhibit heightened sensitivity to user interest drift, as evidenced by their PDM values consistently approximating 0.1 (with the exception of the hashGNN model on the Amazon-books dataset). This observation underscores two critical points: firstly, it highlights the necessity for more frequent model updates in traditional systems to maintain relevance. Secondly, and perhaps more intriguingly, it accentuates the robustness of LLMs in mitigating the effects of shifting user interests, which suggests that LLMs could potentially offer a viable solution to the user interest drift problem. 
    \item When comparing two LLM-based baseline methods, we observed that TALLRec, which uses direct instruction tuning, has comparable performance with the best traditional methods (HashGNN). However, BinLLM, which incorporates collaborative information, demonstrated significant advantages across all scenarios. This underscores the necessity of personalizing the prompt for understanding the user interests.
    \item When comparing the two datasets, we observed that baseline models were more significantly impacted by the Amazon-Movie dataset than the Amazon-Books dataset. This was particularly evident in the HashGNN model, where the PDM increased from 0.0285 to 0.1100. This discrepancy may be attributed to the Movie dataset's broader time span, which resulted in more dramatic changes in user interests between the training set ($D_{train}$) and the test set ($D_{9}$).These findings further emphasize the importance of addressing user interest drift in recommendation systems.
    \item When comparing the performance of BinLLM and BinLLM+, we observed that BinLLM+ significantly underperformed BinLLM. This finding suggests that although BinLLM can improve its performance via collaborative information, it is optimized via training for a specific collaborative model. Consequently, simply updating the collaborative model leads to a decline in model performance due to overfitting on this specific collaborative model.  
    \item When comparing RecICL-TALLRec and RecICL-BinLLM, we observe that their performance is remarkably similar. Contrary to expectations, the advantage of BinLLM's use of collaborative information is not clearly evident within the RecICL framework. This unexpected outcome may be attributed to two factors: (1) The global collaborative information provided by the collaborative model may not accurately reflect user interests as effectively as the user's most recent interactions. (2) There might be an ongoing issue with the collaborative model's performance degradation. Despite these observations, it's important to note that when comparing RecICL-BinLLM with the standalone BinLLM, we still see a significant performance improvement. This contrast underscores the effectiveness and high adaptability of our proposed RecICL method.

\end{itemize}

Next, we will further analyze the RecICL method, our subsequent experiments will be based on \zjz{RecICL-{TALLRec}} since it shows the best performance in our main experiment.

\subsection{Robust Analysis (RQ2)}
\begin{figure}[t]
        \includegraphics[width=0.46\textwidth]{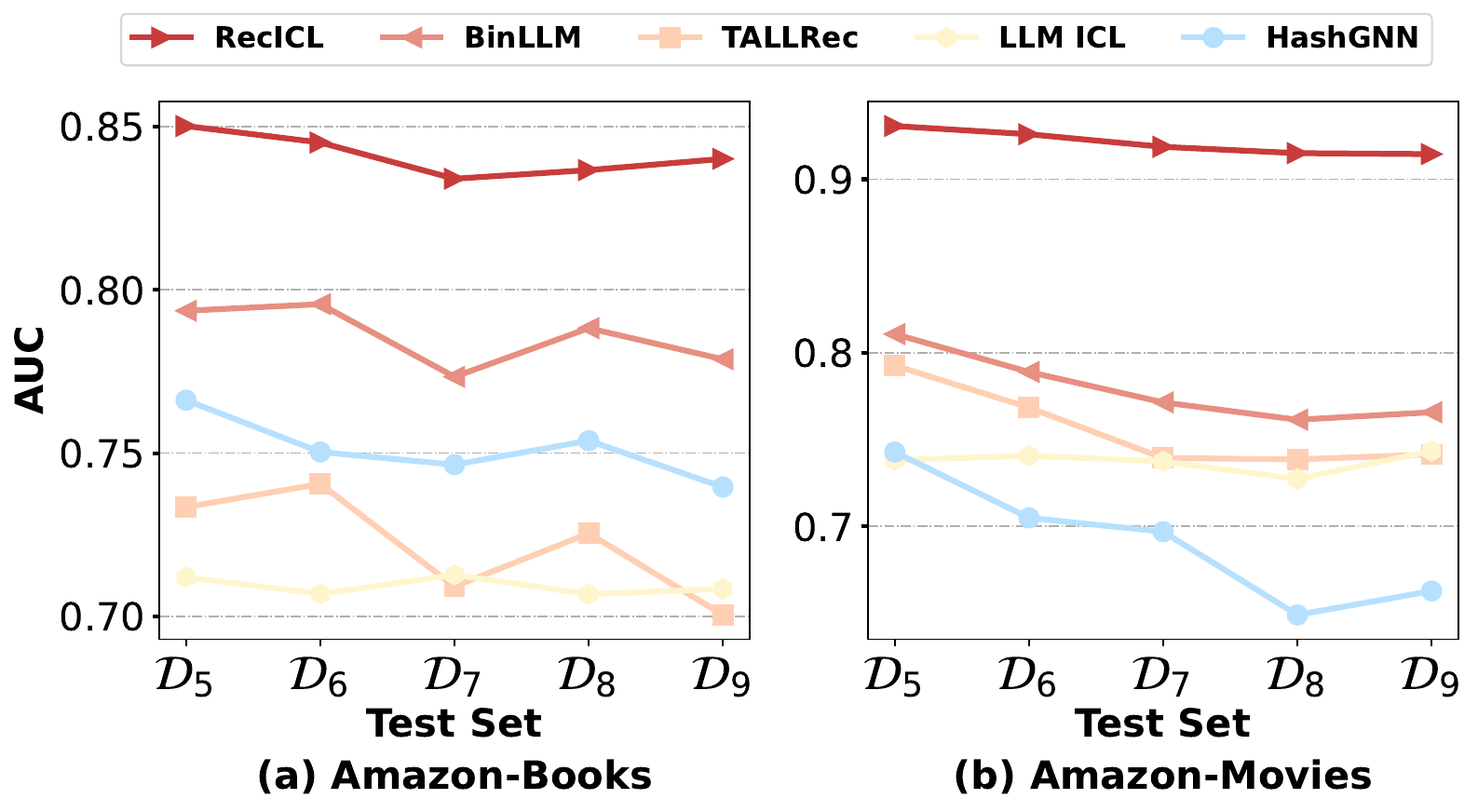}
    \vspace{-1em}
    \caption{The performance of the model on different test sets after training on $\mathcal{D}_{train}$. The x-axis represents practical data partitions, with larger subscripts indicating a greater shift in user interests compared to the training set. The y-axis shows the corresponding AUC metric for each data partition.}
    \vspace{-0.4cm}
        \label{fig:time_decay}

\end{figure}

To show the robustness of RecICL, validate its performance across different time periods, we present the results of several LLM-based \zjz{methods} trained on $D_{train}$ and evaluated on datasets $D_{5},D_{6},D_{7}, D_{8},$ and $D_{9}$ in Figure~\ref{fig:time_decay}, respectively. The main findings are as follows:
\begin{itemize}[leftmargin=*]
    \item In terms of comparative performance, RecICL demonstrates consistent advantages across all periods, further validating the effectiveness of our approach. Even more encouraging is that when tested on $D_{5}$, where user interests have not undergone dramatic changes, our method still shows significant improvements over baseline approaches. We attribute this to the fact that while employing in-context learning, we essentially provide more personalized user input (each user's most recent interaction and its feedback), enabling the model to better model the user. This effectively personalizes the input prompt, leading to substantial improvements. 
    \item When comparing all other LLM-based methods, we found that ICL performance remains relatively stable. Although ICL yields the lowest performance, it does not overfit user preferences from specific periods due to the absence of domain-specific fine-tuning. This further explains the stable performance of RecICL. 
    \item Furthermore, by examining the performance of HashGNN and BinLLM in both datasets, we observe that BinLLM is more susceptible to the influence of the collaborative models  it relies on, i.e., HashGNN. When the collaborative models are significantly affected by shifts in user interests, BinLLM is also substantially impacted. Nevertheless, BinLLM still demonstrates greater stability compared to HashGNN, underscoring the robustness of LLM-based approaches.
\end{itemize}

\subsection{In-depth Analysis (RQ3)}
Next, we will conduct a more in-depth analysis of RecICL.
\subsubsection{Influnce of Few-shot Number}

\begin{figure}[t]
        \centering
        \includegraphics[width=0.5\textwidth]{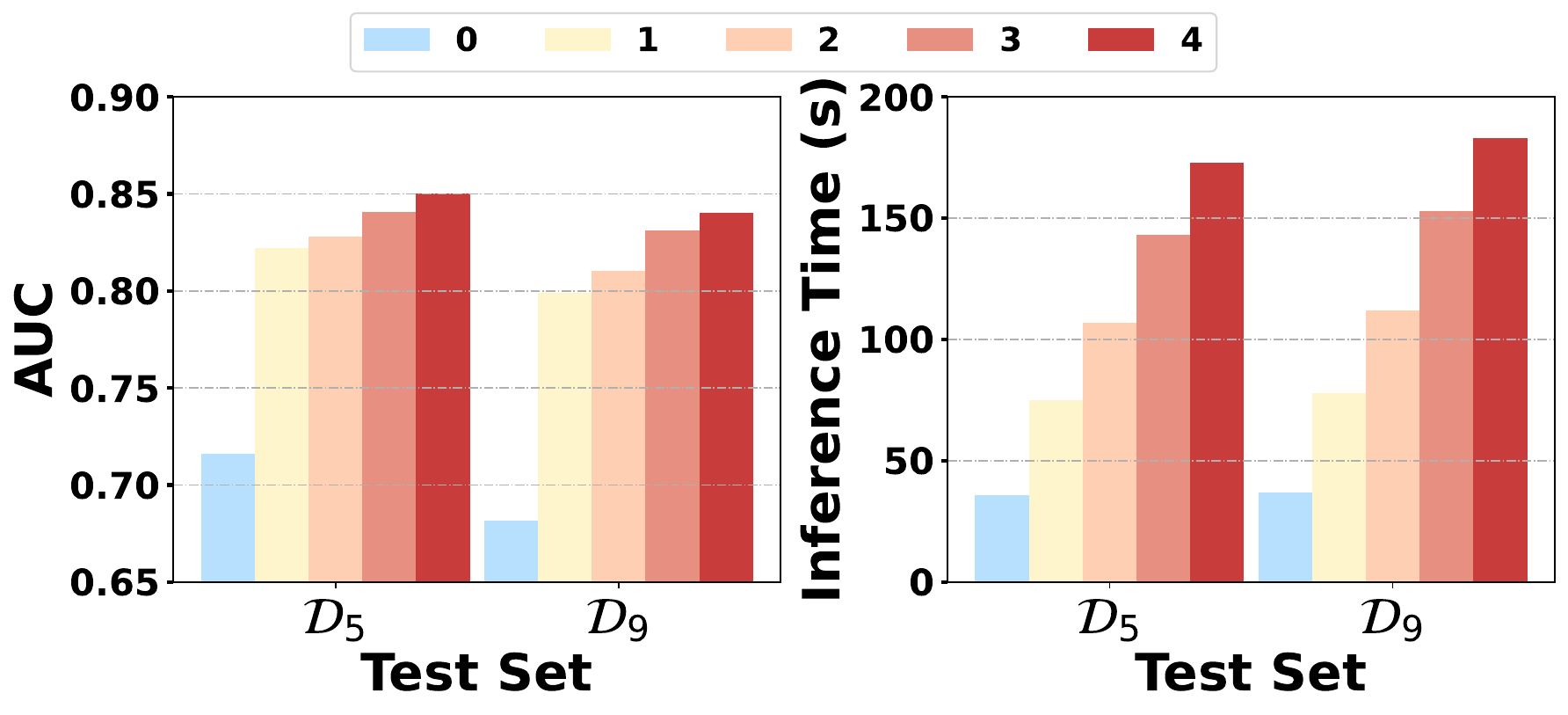}
    \vspace{-2em}
    \caption{The performance of  RecICL trained with varying numbers of few-shot samples (left) and its inference overhead on the entire test set (right). Note that when the number of few-shot samples is 0, it is equivalent to TALLRec.}
    \label{fig:few-shot-number}
    \vspace{-1em}
\end{figure}

We first analyze the number of few-shot samples when applying RecICL, as shown in Figure~\ref{fig:few-shot-number}. The figure illustrates the model's performance on $D_{5}$ and $D_{9}$ and the inference time changes for all 5000 samples. We can draw the following conclusions:
\begin{itemize}[leftmargin=*]
    \item When considering the overall performance, regardless of the number of few-shot samples chosen, the model's performance shows a qualitative improvement compared to zero samples (i.e., TALLRec). When considering Table~\ref{table:main_result}, even RecICL with just one few-shot sample demonstrates a clear performance advantage over the previous SOTA method, BinLLM, which further demonstrates the effectiveness of our approach.
    \item Besides, we found that as the number of few-shot samples increases, the model continues to improve its recommendation performance. However, the most significant performance boost occurs when the number of few-shot samples increases from 0 to 1. This further indicate the importance of user's most recent interaction which directly reflect their current preference.
    \item When looking at the right figure, we observed that the inference time grows approximately linearly with the increase in few-shot samples, mainly due to the increased input length. This issue could potentially be addressed through prefill optimization.
\end{itemize}
In summary, there is a trade-off between the performance gains and the increased inference time brought by few-shot samples. When prioritizing performance, more few-shot samples can be used; when seeking balance, using 1 or 2 few-shot samples can bring noticeable performance improvements.

\subsubsection{Impact of Few-shot Selection}
\begin{figure}[t]
    \centering
    \includegraphics[width=0.45\textwidth]{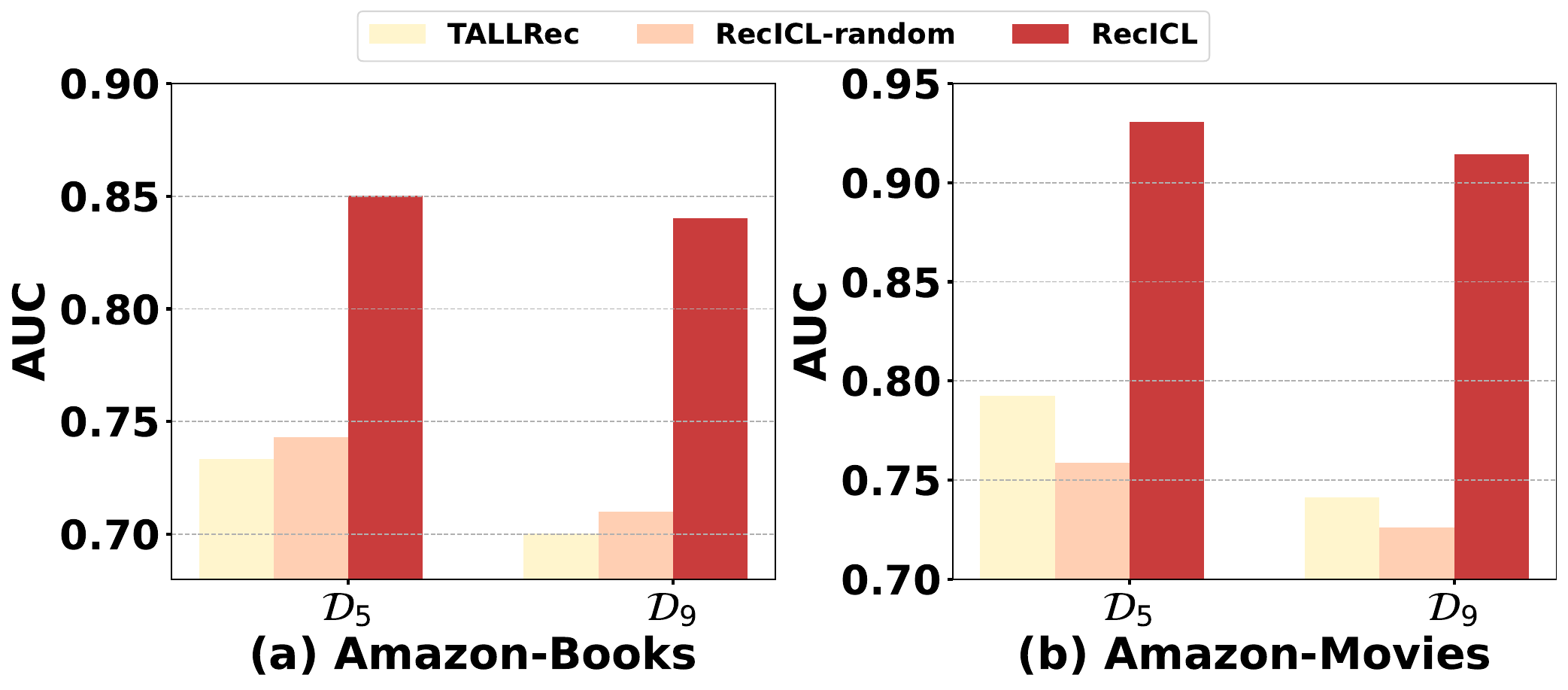}  
    \caption{Performance comparison of TALLRec, RecICL, and RecICL using random select few-shot samples on $D_{5}$ and $D_{9}$.}
    \label{fig:ablation_selection}  
\end{figure}

%
Now we delve into the few-shot selection strategy of RecICL, and aim to answer the following question: How impactful is leveraging users' recent interactions and feedback?  
To answer this question, we fist present a ablation study, which compare the performance ammong TALLRec, $RecICL-random$ using four randomly selected samples as few-shot examples, and $RecICL$.
As shown in the figure~\ref{fig:ablation_selection}, we observe that with random samples, $RecICL-random$ does not always show performance improvement compared with TALLRec, demonstrating the importance of choosing the most recent interaction as few-shot samples.  This further verified that when using RecICL, we need to perform instance-level personalization to capture the user's real-time interest thereby improving the recommendation accuracy.



\begin{figure}[t]
    \centering
    \includegraphics[width=0.45\textwidth]{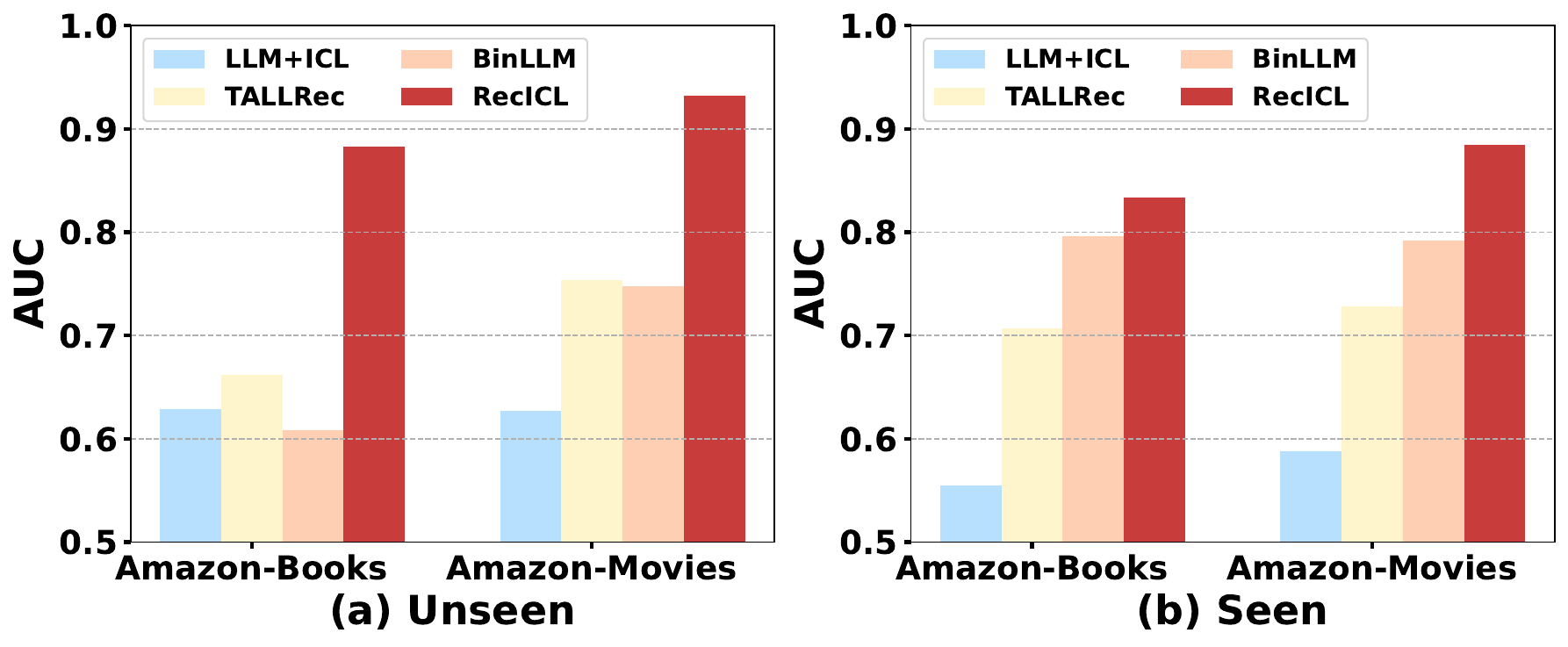}  
    \caption{Performance comparison on seen (right) and unseen (left) users on Amazon-Books and Amazon-Movies datasets.}
    \vspace{-1em}
    \label{fig:seen_unseen}  
\end{figure}

Apart from the ablation study, we also consider the most extreme scenario of user interest drift is when models have never encountered a user during thier training phase. Consequently, this user's interests are entirely unknown to the model, and we can only learn about the user's preferences through their real-time feedback. To validate that RecICL is also effective on this scenario, we divided the interactions in the test set into two categories based on whether the user has appeared in the training set. We then calculated the recommendation performance for each category separately. The results are illustrated in Figure~\ref{fig:seen_unseen}.

Our findings indicate that RecICL demonstrates significant performance improvements compared to the baseline for users not present in the training set. This can be attributed to RecICL's ability to effectively leverage recent interactions from new users to model their interests. Besides, for RecICL, we also observed that in addition to smaller performance gains among user groups previously encountered by the model, the overall performance was also slightly lower compared to unseen users. This phenomenon may be caused by the fact that new users tend to have more focused interest preferences closely related to their recent interactions, while old users might have more complex, long-term interests that are not fully captured by our in-context learning input.
\section{Conclusion and Future Work}


In this paper, we highlight the challenges faced by LLMs in recommender systems when dealing with user interest drift. Unlike traditional models, LLMs cannot timely update their parameters due to high training costs, even with current acceleration techniques. To address this issue, we propose RecICL, which ensures that the LLM aligns with the recommendation scenario while preserving and enhancing its in-context learning capabilities in the recommendation context. During deployment, it can utilize the user's most recent feedback by inputting this feedback as few-shot samples to the model, allowing it to capture the user's real-time interests. Extensive experimental results also illustrate the effectiveness and adaptability of RecICL, successfully making real-time recommendations without any model-level updates.
In the future, we aim to delve deeper into this research direction. On one hand, we'd like to explore ways to enable LLMs to better utilize collaborative information from updated traditional models, aligning with existing incremental learning methods. On the other hand, we consider incorporating users' long-term interests into the in-context learning inference process to further enhance model performance. 

\begin{acks}
acknowledgments
\end{acks}

\bibliographystyle{ACM-Reference-Format}
\bibliography{sample-base}

\appendix

\end{document}